\documentstyle[aps,prl,epsf,floats]{revtex}
\advance\topmargin by -.1in
\advance\textheight by 0.15in
\begin{document}

\twocolumn[\hsize\textwidth\columnwidth\hsize\csname
@twocolumnfalse\endcsname

\title{Instability of charge ordered states in doped
  antiferromagnets} 

\author{Leonid P. Pryadko$^*$,  Steven Kivelson$^*$ and Daniel
  W. Hone$^\dagger$} 

\address{$^*$Department of Physics \& Astronomy, 
  University of California, Los Angeles, CA  90095}
\address{$^\dagger$Institute for Theoretical Physics, University of
California, Santa Barbara, CA 93106}
\date\today
\maketitle
\begin{abstract}
  We analyze the induced interactions between localized holes in {\it
    weakly-doped} Heisenberg antiferromagnets due to the modification
  of the quantum zero point spin wave energy; {\it i.e.} the analogue
  of the Casimir effect.  We show that this interaction is {\em
    uniformly attractive\/} and falls off as $r^{-2 d+1}$ in $d$
  dimensions.  For ``stripes'', {\it i.e} parallel $(d-1)$-dimensional
  hypersurfaces of localized holes, the interaction energy per unit
  hyperarea is attractive and falls, generically, like $r^{- d}$.  We
  argue that, in the absence of a long-range Coulomb repulsion between
  holes, this interaction leads to an instability of {\it any}
  charge-ordered state in the dilute doping limit.
\end{abstract}
\vskip0.5pc]
\narrowtext

It is still not clear what happens when a dilute concentration of
holes is introduced into a quantum Heisenberg antiferromagnet (AF).
This is one of the central issues in the theory of correlated
electronic systems, especially as it relates\cite{review} to the high
temperature cuprate superconductors and related oxides.  One class of
proposals\cite{Zaanen-89,LosAlamos,White-97} holds that the result is
a spatially inhomogeneous ``charge ordered'' ground state.
Unfortunately, numerical analysis of the stability of such states is
often inconclusive because the typical energy differences between
states are small, and the Goldstone modes (spin-waves) produce finite
size effects which decrease slowly with system size.

\begin{figure}[htbp]
  \def\ifundefineddddd#1{\expandafter\ifx\csname#1\endcsname\relax}%
  \ifundefineddddd{epsfbox}\relax\else%
  \epsfxsize=0.75\columnwidth%
  \strut\hfill\epsfbox{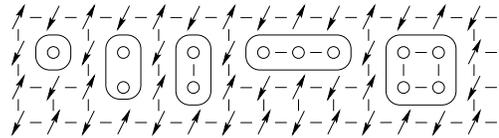}\hfill\strut\vskip3pt
  \fi%
  \caption{Clusters of localized holes are modeled by removing
    the bonds 
    connecting the spins to the rest of the system.  The additional
    bonds inside the disconnected areas do not contribute to the
    interaction between hole clusters.}
  \label{fig:areas}
\end{figure}

The goal of this paper is to investigate the thermodynamic stability
of charge-ordered states in short-range AF spin models in the dilute
doping limit.  We calculate the induced interaction between well-separated
clusters of
holes due to their modification of the spin wave spectrum, and find
that it is {\em uniformly attractive}.  Specifically, the asymptotic
long distance ($r\!\rightarrow\!\infty$) interaction between two
hole clusters (see Fig.~\ref{fig:areas}) is of the form
${\cal E}\sim -JS\, r^{-2 d+1}$ for a $d$-dimensional, spin-$S$
Heisenberg quantum AF with exchange coupling $J$ (see
Eq.~\ref{eq:afm-spins}).  For extended clusters of holes, under
generic circumstances, the dependence on separation (but not the
absolute magnitude) of the interaction energy can be reliably
estimated by summing the pairwise hole-hole interaction over all
pairs; for example, the interaction per unit hyperarea between
parallel walls of localized holes ({\it i.e.}  codimension $1$
hypersurfaces which, with the case of $d\!=\!2$ in mind, we refer to
as ``stripes'') falls with their separation as ${\cal E}\!\sim\!
-JS\, r^{-d}$.

As a consequence of this attraction, in the absence of a long-range
Coulomb repulsion between holes, all charge-ordered states with
sufficiently small hole concentration are unstable to phase
separation, although it is possible that there exist non-vanishing
hole densities for which charge ordered states are
stable\cite{LosAlamos,White-97}.  Remarkably, we find that the
correct asymptotic form of the induced interactions cannot be
obtained in any finite order of na\"\i{}ve perturbation theory,
because of the singular effect of a marginally bound (zero-energy)
spin-wave state associated with the fact that doped holes actually
change the Hilbert space, by changing the number of spins.  However, a
simple modified perturbation theory can be constructed which
qualitatively reproduces the exact spin-wave results in all cases we
have tested.

\begin{figure}[htbp]
  \def\ifundefineddddd#1{\expandafter\ifx\csname#1\endcsname\relax}%
  \ifundefineddddd{epsfbox}\relax\else%
  \epsfxsize=0.75\columnwidth%
  \strut\hfill\epsfbox{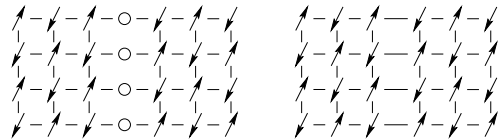}\hfill\strut\vskip3pt
  \fi%
  \caption{A stripe of holes forming an anti-phase domain wall is modeled
    as a line of weak bonds in a perfect AF.}
  \label{fig:strip}
\end{figure}

{\bf The Model:}  The number of spin degrees of freedom changes with doping
and, therefore, the Hilbert spaces appropriate to the doped and
undoped AF are different.  Therefore, for mathematical convenience it
is preferable to use a
model with a spin $S$ operator on each site, and to treat any system
with localized holes as a limiting case in which the coupling between
a set of ``impurity'' sites and its neighbors goes to zero.  For
static holes, this is all there is to the model.  However, so long as
the holes are {\it localized}, either by an impurity potential or by a
self-consistent field (as in Hartree-Fock
solutions\cite{Zaanen-89}) the effect of hole-hopping can be
treated by including a larger set of modified exchange interactions in
the neighborhood of each hole. Thus, the spin Hamiltonian of the doped
system differs from that of the pure AF only in the strength of some
exchange couplings:
\begin{equation}
  \hat H=\hat H_0-\lambda \hat H'\equiv\hat H_0-\hat V,
  \label{eq:general-perturb}
\end{equation}
where $H_{0}$ is the Hamiltonian for the perfect antiferromagnet,
which, for concreteness, we take to have only nearest-neighbor
interactions on a hypercubic lattice,
\begin{equation}
  \label{eq:afm-spins}
  \hat H_0={\textstyle\sum_{\langle ij \rangle}}\,\hat H_{ij},\quad
  \hat H_{ij}=J\, \hat{\bf S}_i\cdot \hat{\bf S}_j,
\end{equation}
and the perturbation Hamiltonian $\hat H'$ 
specifies a set of pairwise exchange interactions such that, in the
limit $\lambda\!=\!1$, a spin near which the hole is localized is
disconnected from the rest of the system.  Clearly, the {\em
  interaction\/} energy between hole clusters is obtained correctly in
this limit, although the cluster self-energy could depend on the
interactions between the fictitious, disconnected spins.
We start with the simplest case of strongly localized holes, for which
$H'$ consists of the sum over the exchange interactions connecting the
designated hole sites in a cluster to the nearest-neighbor sites
surrounding it.  One exceptional geometry which we treat differently
is a stripe which is simultaneously an anti-phase domain wall in the
AF order.  Such a stripe can be treated\cite{Castro-96}, as shown in
Fig.~\ref{fig:strip}, effectively as a wall of {\em bonds\/} with
altered exchange coupling, so that we work in the proper Hilbert space
from the beginning; here na\"\i{}ve perturbation theory in powers of
$\lambda$ yields qualitatively correct generic results.

{\bf Casimir Energy of Stripes:} We begin by considering the
interaction energy between two stripes of static holes; this turns out
to be the simplest problem because for $\lambda\!=\!1$ the region
between the two stripes is cut off from its surroundings, and,
according to a very general argument originally due to
Casimir\cite{Casimir-48}, the interaction energy must fall off as ${\cal
  E}\!\sim\!-r^{-d}$. To be more explicit, if we take the stripes to
be perfectly reflecting, the perpendicular component of the
spin-wave's momentum in the region of width $r$ between the stripes is
discretely quantized, which modifies its vacuum energy.  Then the
distance-dependent part of the energy per unit hyperarea can be
expressed as the difference
$$
{\cal E}(r)=\!\sum_{k_{\perp}=\pi n/r}\!\int\!
{d^{d-1} k_{\parallel}\over  (2\pi)^{d-1}}
\left ({\hbar\omega_{\vec k}\over 2}\right )-r\, E_{\infty},
$$
where $\hbar \omega_{\vec k} $ is the spin-wave energy, and 
$E_{\infty}$ is the vacuum energy density of the infinite lattice.
The summation, performed with the Poisson formula, gives
\begin{eqnarray}
  {\cal E}&=&r \sum_{m\neq0}\int {d^dk\over(2\pi)^d}
  {\hbar\omega_{\vec k}\over2} e^{2 i k_\perp r m}\nonumber \\
  \label{eq:univers}
  &=&-{c\over r^d}\,\left[{\zeta(d\!+\!1)\,\Gamma(d\!+\!1)
    \over2^{2d}\,\pi^{d/2}\,\Gamma(d/2)}\right]
  +{\cal O}(r^{-d-1}),
\end{eqnarray} 
which is proportional to the spin-wave velocity, $c$ ($\hbar
\omega_{\vec k}\approx c |\vec k|$; in linear spin-wave
approximation\cite{Auerbach-Book} $c\!=\!JS/\sqrt d$,) since only small
momenta contribute to the large-dis\-tance asymptotics of the
interaction energy.  This result is universal: 
it depends on the number of acoustic modes and their speed but not on
the form of the spectrum at high energies or the specific boundary
conditions.

{\bf Linear Spin-Wave (LSW) Theory:} %
In more general circumstances we need to compute the interactions from
a more microscopic approach.  For this purpose, we adopt LSW theory
which is quantitatively accurate for large $S$, but which we expect to
be a reliable method for extracting the long distance physics for
$d\!\ge\! 2$ even for $S\!=\!1/2$, since already in $d\!=\!2$ AF order
is very robust\cite{CHN-89}.  In order to make contact with the
perturbative results discussed below, we calculate the LSW
correction to the ground-state energy in the presence of localized
stripes or holes at arbitrary $\lambda$.  The exact ground-state
energy can be calculated as a coupling constant integral of the
expectation value of the perturbation Hamiltonian using the
Feynman-Hellman formula, or, in the LSW approximation, expressed as
the functional determinant
\begin{equation}
  \label{eq:fluct-determ-integration}
  -{\cal E}(\lambda)={
    {1\over2}} {\rm Tr\,}\left(\ln\hat{\cal G}^\lambda\!-\ln\hat{\cal
      G}^0\right).
\end{equation}
The latter can be rewritten as a coupling-constant integral of the
diagonal part of the exact LSW Green's function (GF) $\hat{\cal
  G}^\lambda$, as we shall see below.

{\it LSW theory for stripes:\/} As a first application of LSW theory,
we recompute the interaction energy between two stripes oriented along
the $y$ axis in $d\!=\!2$.  In an obvious mixed representation,
labelled by the conserved wave vector $k_y$ parallel to the stripes
and the lattice position $x$ in the direction perpendicular to the
stripes, 
the Dyson equation for the GF reduces to a finite sum,
\begin{equation}
  \label{eq:dyson-stripe}
  \hat{\cal G}^\lambda_{x,x'}(\omega,k_y)\!=\! %
  \hat{\cal G}^{0}_{x\!-\!x'}(\omega,k_y)\!+\! %
  \lambda\,{\textstyle \sum_i}\,\hat{\cal G}^{0}_{x\!-\!x_i}\,
  \hat{\cal G}^\lambda_{x_i,x'},
\end{equation}
where $i$ labels the vertical lines of sites connected by the weak
bond representing the effect of the stripes,
and the unperturbed GFs are given explicitly by
\begin{equation}
  \label{eq:sw-gf}
  \hat{\cal G}^{0}_{{\bf r}}\!=\! \left|
    \begin{array}[c]{cc}
      {\cal D}_{\bf r}&{\cal F}_{\bf r}\\
      \bar{\cal F}_{\bf r}&\bar{\cal D}_{\bf r}
    \end{array}
  \right|\!=\!
  \left|
    \begin{array}[c]{cc}
      \langle T_\tau s_{\bf r}(\tau)\bar s_{0}(0)\rangle &
      \langle T_\tau s_{\bf r}(\tau) s_{0}(0)\rangle  \\
      \langle T_\tau\bar s_{\bf r}(\tau)\bar s_{0}(0)\rangle&
      \langle T_\tau\bar s_{\bf r}(\tau) s_{0}(0)\rangle
    \end{array}
  \right|,
\end{equation}
where the operators $s_{\bf r}\!=\!b_{\bf r}\!+\!b^\dagger_{{\bf
    r}+\hat{\bf x}}$, $\bar s_{\bf r}\!=\!b^\dagger_{\bf
  r}\!+\!b_{{\bf r}+\hat{\bf x}}$ are defined on the bonds $\langle
{\bf r},{\bf r}\!+\!\hat{\bf x}\rangle$ in terms of the
Holstein-Primakoff boson operators $b_{\bf r}$, $b^\dagger_{\bf r}$.
As for the problem of a quantum particle in the presence of a finite
number of point scatterers, the solution of
Eq.~(\ref{eq:dyson-stripe}) involves the inversion of only a finite
matrix, and the functional
determinant~(\ref{eq:fluct-determ-integration}) can be rewritten in
the convenient form,
\begin{equation}
  \label{eq:fluct-determ-simplified}
  {\cal E} =-{J\,S\over2}\!\!\int_0^\lambda\!{d\lambda'\over\lambda'}
  \!\!\int\!{dk_y d\omega\over4\pi^2}\sum_i\left(\hat{\cal
      G}^{\lambda'}\!-\hat{\cal G}^{0}\right)_{ii},
\end{equation}
with integrations over the coupling constant $\lambda$, the
zero-temperature Matsubara frequency $\omega$, and the conserved
momentum $k_y$ along the stripes, along with the finite summation over
the stripe index $i$.

The matrix elements of the pure crystal GF (\ref{eq:sw-gf})
in the mixed representation are given explicitly  by the integrals
\begin{eqnarray}\nonumber
  {\cal D}_x(\omega,k_y) \!
  &=&\!\!   
  \int\! {dk_x\over4\pi} {1\!-\!\gamma_k\cos k_x\over
    \omega^2+\epsilon_k^2}\,e^{i\,k_x x},\\
  \label{eq:stripes-matr-ss}
  {\cal F}_x(\omega,k_y)\! &=&
   \!\! \int\! {dk_x\over4\pi} {\cos k_x\!-\!\gamma_k\!-\!\omega\sin k_x\over
    \omega^2+\epsilon_k^2}\,e^{i\,k_x x},\\\nonumber
  \bar {\cal D}_x(\omega,k_y)\! &=&\! {\cal D}_x(-\omega,k_y),
  \quad
  \bar{{\cal F}}_x(\omega,k_y)\!=\!{\cal F}_x(-\omega,k_y),
\end{eqnarray}
where $\gamma_k=[\cos(k_x)+\cos(k_y)]/2$ and
$\epsilon_k^2=1-\gamma_k^2$.  For the case of only two anti-phase
stripes separated by the distance $r\!=\!|x_2\!-\!x_1|$, the
cou\-pling-con\-stant integration~(\ref{eq:fluct-determ-simplified})
can be performed analytically, with the result
\begin{eqnarray}
  \label{eq:tr-log-answer}
  {{\cal E}\over J\,S}\!=\!\!  \int\!\!{d\omega\,dk_y\over8\pi^2}\!\!
  \sum_{\eta=\pm 1}\!\!\ln{ |\lambda^{-1}\!\!-\!{\cal D}_0\!+\!\eta
    {\cal D}_r|^2\!\!-\!|{\cal F}_0 \!-\!\eta {\cal F}_r|^2\over
    |\lambda^{-1}\!-\! {\cal D}_0|^2\!-\!|{\cal F}_0|^2}.
\end{eqnarray}
The matrix elements of the bare spin-wave GF~(\ref{eq:sw-gf}) are
small at $r\rightarrow\infty$, 
so {\it for all\/} $\lambda\!<\!1$, an asymptotic expression for the 
interaction energy can be obtained by expanding
Eq.~(\ref{eq:tr-log-answer}) in powers of ${\cal F}_r$, ${\cal D}_r$;
to leading order we obtain
\begin{equation}
  {\cal E} = - {JS\,\lambda^2\over r^4}
  \left[\frac{240\lambda(1\!-\!\lambda)\!  +\!3\sqrt2 \left( 51\!  -\!
      102\lambda\! +\! 67\lambda^2 \right) }{1024\,\pi ( 1\!
    -\!\lambda)^2 {{\left[ 2\! -\!  \left( 2\! -\! {\sqrt{2}} \right)
            \lambda \right] }^2}}\right].
  \label{eq:lt1}
\end{equation}
The na\"\i{}ve perturbative results for this energy can be obtained by
further expanding this expression to second order in $\lambda$:
\begin{equation}
  \label{eq:strip-pert-energy}
  {\cal E}^{\lambda\rightarrow0}(r)=-{JS\,\lambda^2\over r^4}\,
  \left[{153\over2048\,\pi\sqrt2 }\right]+{\cal O}\left(r^{-5}\right).
\end{equation}
At $\lambda\!=\!1$, however, the expression~(\ref{eq:lt1}) diverges,
because the denominator of the argument of the logarithm in
Eq.~(\ref{eq:tr-log-answer}) has a zero at the point
$\omega\!=\!k_y\!=\!0$.  It is this zero, identified as the
zero-energy spin wave state bound to each stripe, which is ultimately
responsible for the modification of the asymptotic form of the
interaction energy.  As discussed later, the existence of such a state
is related to the 
change in the structure of the Hilbert space as the clusters become
isolated at this value of $\lambda$; such a state exists near a hole
cluster of any geometry, and it cannot be eliminated by corrections
due to spin-wave interactions or to the holes' mobility.  In our
calculation, we account for this zero-energy state by solving the
scattering problem near each stripe or hole cluster exactly.

\begin{figure}[tbp] %
  \def\ifundefineddddd#1{\expandafter\ifx\csname#1\endcsname\relax}%
  \ifundefineddddd{epsfbox}\relax\else%
  \epsfxsize=\the\columnwidth%
  \epsfbox{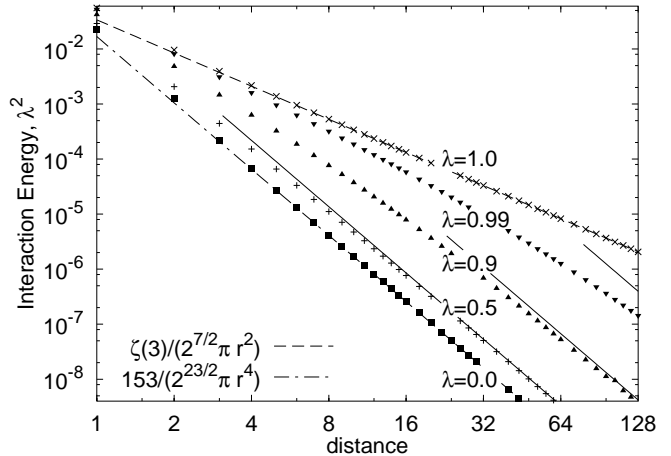}\vskip3pt
  \fi%
  \caption{The LSW Casimir energy of two stripes in units of $J S^2
    (\lambda^2/S)$ from
 Eq. (9).  The $\lambda=0$ line indicates the perturbative
    result.  The solid lines show the calculated asymptotic behavior
    at $r\rightarrow\infty$ at intermediate values of $\lambda$.}
  \label{fig:striplots}
\end{figure}

The correct asymptotic behavior at $\lambda\!=\!1$ can be ob\-tained
by re-evaluating the interaction energy, starting with the complete
expression~(\ref{eq:tr-log-answer}).  Using the long-distance
asymptotics of the components~(\ref{eq:stripes-matr-ss}) of the GF, we
obtain to leading order in $1/r$
\begin{displaymath}
  {\cal E}^{\lambda=1}(r)={JS\over2 \,r^2}\int {dk_y\,d\omega\over(2\pi)^2} 
  \ln\left[1-e^{-2\,(k_y^2+2\omega^2)^{1/2}} \right],
\end{displaymath}
which leads to the universal Casimir result~(\ref{eq:univers})
evaluated at $d\!=\!2$ and $c=JS/\sqrt2$.  We have also investigated
the crossover from the perturbative
expressions~(\ref{eq:strip-pert-energy}), (\ref{eq:lt1}) at small
values of $\lambda$ to the universal form~(\ref{eq:univers}) at
$\lambda\!=\!1$ by integrating the exact LSW
energy~(\ref{eq:tr-log-answer}) numerically.  The results are shown in
Fig.~\ref{fig:striplots}, along with the corresponding asymptotic
expressions.

{\it LSW Theory for Isolated Holes:\/} Zero-energy spin-wave bound
states and the associated divergence of the perturbation series happen
not only for extended objects like the stripes we just considered, but
also for solitary holes or bigger hole clusters.  As a second example,
we consider explicitly the interactions between two isolated holes.
The Hamiltonian of a single hole can be expressed as a sum of terms
with different symmetries with respect to the point
group\cite{Bulut-89}, and we find that in arbitrary dimension, a
zero-energy, totally symmetric bound-state emerges at the same
critical value $\lambda\!=\!1$, so that the corresponding spin-wave
scattering amplitude diverges at small frequencies.  Thus, while for
any $\lambda\!<\!1$, the interaction between two holes has the same
asymptotic behavior as the leading order perturbative expression,
\begin{equation}
  \label{eq:two-spins-d-dimens}
  {\cal E}_{\rm pert}=-{JS\,\lambda^2\over r^{2d+1}}\, 
  \left[{( d^2\!-\!1)\,( d\!+\!1)\,\Gamma({d\!+\!\frac{1}{2}}) \, \Gamma({d
      + 1\over2}) \over 2^{d -
      1}\,{\pi^d}\,{d^{{{3}/{2}}}}\,(d\!+\!2)\,
    \Gamma({\frac{d}{2}})}\right],
\end{equation}
for $\lambda\!=\!1$ the asymptotic form of the interaction is
\begin{equation}
  \label{eq:gf-resonant-bare}
  {\cal E}^{\lambda=1}=-{JS\over r^{2d-1}}\left[ \frac{
    \Gamma(d-{\frac{1}{2}} )\, \Gamma({\frac{d + 1}{2}})}{2^{d}{{\pi
        }^d}\,{\sqrt{d}}\, \Gamma({\frac{d}{2}})}\right],
\end{equation}
which is valid both for holes at odd separations (different
sublattices, holes with opposite spins) and holes at even
separations (same sublattice, holes with the same spin).  

{\it Marginally-Bound States and the Proper Way to Do Perturbation
  Theory:\/} We return now to the issue of the failure of na\"\i{}ve
perturbation expansion in powers of $\lambda$.  For a hole cluster of
arbitrary 
geometry the Dyson equation [Eq.~(\ref{eq:dyson-stripe}) for the case
of stripes] gives bound states when
\begin{equation}
  \label{eq:sw-bound}
  \det \left|\hat {\cal G}^0_{{\bf r}_i-{\bf r}_j}(\omega)
    -\delta_{ij}\lambda^{-1}\right|=0,
\end{equation}
where the two-component GF~(\ref{eq:sw-gf}) is calculated for each
pair of bonds present in the perturbation Hamiltonian $\hat H'$ as
specified in Eq.~(\ref{eq:general-perturb}).  At $\lambda\!=\!1$, where
the artificially introduced spins become disconnected, they acquire the
freedom to rotate with respect to the rest of the system, which
reveals itself as a zero-energy spin-wave state localized on these
spins.  Because of the continuity of the GF in
Eq.~(\ref{eq:sw-bound}), this implies the existence of a soft
spin-wave mode coupled to the holes, which is responsible for the
divergence of the spin-wave scattering amplitudes off the hole cluster
at small frequencies, and also provides the singularity required to
modify the Casimir interaction energy obtained perturbatively.

We checked this argument by computing the eigenvalues of the matrix
$\hat1\!-\!\hat{\cal G}^0_{{\bf r}_i-{\bf r}_j}(\omega\!=\!0)$ for
several hole cluster geometries, including those shown in
Fig.~\ref{fig:areas}.  In agreement with the prediction, there are
exactly two zero eigenvalues for the clusters with only one
disconnected group of spins, one for each spin wave branch, and there
are four such eigenvalues for the geometry in the second example in
Fig.~\ref{fig:areas}, where the two spins of the cluster are
disconnected from each other.  All other eigenvalues are positive,
implying that the perturbation expansion is free of singularities for
$\lambda\!<\!1$.  Within the same approach, we have also considered
the effect of virtual hops of less strongly localized holes, 
which induce
additional AF exchange couplings between the
spins in the neighborhood of the hole.  Once again, the zero-energy
spin-wave bound state persists regardless of the strengths of these
additional interactions.

Because the existence of a zero-energy spin wave bound state within an
isolated cluster representing localized holes follows from symmetry
arguments, it is a very robust feature of the spectrum.  However, the
degeneracy of the artificially inserted spins can be lifted by a local
magnetic field term, added to the perturbation Hamiltonian $\hat H'$
(obviously, this does not modify the physics of the rest of the
system).  It turns out that this term not only removes the divergence
of the perturbation series, but also contributes an energy ${\cal
  E}\!\sim\! r^{-2 d+1}$ to the interaction between the hole clusters
already in second order of the perturbation expansion, which now gives
the correct qualitative result.  The improved perturbation expansion
can be used to prove that the Casimir interaction between hole
clusters indeed falls off as ${\cal E}\!\sim\! r^{-2 d+1}$ even in the
presence of spin-wave interactions.

The only case where the non-physical spins need not be introduced is
the stripe serving as an anti-phase domain wall, modeled as shown in
Fig.~\ref{fig:strip}.  Although perturbation theory here also breaks
down at $\lambda\!=\!1$ because of the soft spin-wave mode bound to
the stripe, this mode is related to the freedom to rotate the AF
magnetization of the two parts, which become disconnected at this
value of $\lambda$.  Such symmetry is easily destroyed by the holes'
mobility or by spin-wave interaction corrections, and, therefore, the
robust asymptotic form of the interaction energy between the antiphase
stripes is the one given by the perturbation
expansion~(\ref{eq:strip-pert-energy}).  Note, however, that the
general asymptotic behavior ${\cal E}\!\sim\!  -J' S r^{-d}$ is
restored if next-nearest neighbor interactions are included in the
Hamiltonian~(\ref{eq:afm-spins}).

{\bf Implications:} We have considered Casimir interactions between
well separated hole clusters in AFs.  For hole clusters or
stripes in a uniform AF, this energy is uniformly attractive and
generally falls off with distance as ${\cal E}\!\sim\! r^{-2d+1}$ and
${\cal E}\!\sim\! r^{-d}$ respectively.  The interaction is
quantitatively weak; for two holes in the $S=1/2$ AF the interaction
between next-nearest neighbor holes is less than $10^{{-2}}J$.
However, because the interaction falls slowly with distance, it is
important for an analysis of the stability of static charge-ordered
structures in systems lacking long-range Coulomb repulsion.  It has
been conjectured\cite{LosAlamos,Anderson-97} that phase separation
is a ubiquitous feature of lightly doped antiferromagnets, and that
consequently there is always a first-order transition separating the
undoped and doped states.  Evidence in
support\cite{Lin-90} and in
conflict\cite{White-97,Rice-94} with this conjecture has been obtained
from numerical studies of small-size systems.  Phase separation 
has been shown to occur
\cite{vanDongen-95,Carlson-97} 
in the large $d$ limit of the Hubbard and
$t$-$J$ models,
and in the mean-field spiral states of the 
large $N$ $t$-$J$ model\cite{assa}.  The present results offer strong
additional support for the validity of this conjecture. Specifically,
we claim that because of this Casimir-like interaction, any static
ordered state of neutral holes will be thermodynamically unstable with
respect to phase separation at small enough doping.

\begin{thebibliography}{10}
\bibitem{review}  Experiments and theories are reviewed in Ref.\ \CITE{LosAlamos}.
\end{thebibliography}

{\bf Acknowledgments:} We thank A. H. Castro Neto for conversations.
This work was supported in part by NSF grants
DMR93-12606 at UCLA and PHY94-07194 at ITP-UCSB.

\end{document}